\documentclass[a4paper]{article}

\usepackage{INTERSPEECH2021}
\usepackage{multirow}
\usepackage{makecell}
\usepackage{cite}
\usepackage{url}

\title{HMM-Free Encoder Pre-Training for Streaming RNN Transducer}
\name{Lu Huang, Jingyu Sun$^*$\thanks{* The work was done as an intern at ByteDance.}, Yufeng Tang, Junfeng Hou, Jinkun Chen, Jun Zhang, Zejun Ma}
\address{
Speech \& Audio Team, AI Lab, ByteDance Inc., Beijing, China
}
\email{huanglu.thu19@bytedance.com}

\begin{document}
\maketitle

\begin{abstract}
This work describes an encoder pre-training procedure using frame-wise label to improve the training of streaming recurrent neural network transducer (RNN-T) model. Streaming RNN-T trained from scratch usually performs worse than non-streaming RNN-T.
Although it is common to address this issue through pre-training components of RNN-T with other criteria or frame-wise alignment guidance, the alignment is not easily available in end-to-end manner.
In this work, frame-wise alignment, used to pre-train streaming RNN-T's encoder, is generated without using a HMM-based system. Therefore an all-neural framework equipping HMM-free encoder pre-training is constructed. This is achieved by expanding the spikes of CTC model to their left/right blank frames, and two expanding strategies are proposed. To our best knowledge, this is the first work to simulate HMM-based frame-wise label using CTC model for pre-training. Experiments conducted on LibriSpeech and MLS English tasks show the proposed pre-training procedure, compared with random initialization, reduces the WER by relatively 5\%$\sim$11\% and the emission latency by 60 ms. Besides, the method is lexicon-free, so it is friendly to new languages without manually designed lexicon.
\end{abstract}

\section{Introduction}
\label{introduction}
Recently, end-to-end (E2E) automatic speech recognition (ASR) has been widely explored with the development of deep learning~\cite{lecun2015deep,goodfellow2016deep}, and it is gradually replacing the hybrid framework. In the hybrid framework, acoustic model, pronunciation model and language model (LM) are trained separately using different objective functions, while E2E ASR is an all-neural 
framework with a single objective function.  

ASR can be formulated as a sequence-to-sequence problem. Thus, the alignment path between acoustic feature and label needs to be learned. In the hybrid framework, this is achieved with Hidden Markov Model (HMM). Attention-based Encoder-Decoder (AED)~\cite{chan2016listen,bahdanau2016end,watanabe2017hybrid} methods use attention mechanism to learn such alignment path, but streaming ASR is still a challenge due to the full-sequence attention mechanism.
Connectionist Temporal Classification (CTC)~\cite{alex-ctc-2006,graves2014towards} introduces a blank symbol and it models the alignment path by filling blank symbols automatically.
However, CTC assumes that the output of current frame is independent of previous outputs.

RNN-T~\cite{graves2012sequence,alex-rnnt-2013,google-sak-rnnt-2017,hu2020exploring,sougou-rnnt-2019,google-device-rnnt-2019,baidu-rnnt-2017} is designed for streaming E2E ASR with a frame-synchronous decoding strategy, and it improves CTC by introducing a prediction network to abandon the assumption of conditional independence. Therefore, RNN-T is more powerful than CTC and it has been widely explored recently~\cite{chang2019joint,sainath2020streaming,ghodsi2020rnn,zhang2020transformer,li2020towards,kurata2020knowledge,panchapagesan2020efficient}.

However, the training of E2E ASR models is difficult. This is because they must learn all valid alignment paths between acoustic feature and target symbols jointly without any frame-wise supervision. Moreover, in the absence of future context, streaming ASR models using encoder like long short-term memory (LSTM)~\cite{hochreiter1997long, sak2014long} or causal Transformers~\cite{zhang2020transformer,vaswani2017attention} perform even worse than non-streaming models with full-sequence context, as described in \cite{zhang2016highway,zhang2020transformer}.

There are some works trying to improve the performance of streaming models by using knowledge distillation from non-streaming models
~\cite{kurata2020knowledge,kurata2018improved,takashima2019investigation,kurata2019guiding}. 
However, knowledge distillation from non-streaming to streaming CTC models is difficult, because they have different timings of spikes. Therefore, sequence-level~\cite{takashima2019investigation} and chunk-level best-match~\cite{kurata2018improved} knowledge distillation are proposed. Besides, guided training~\cite{kurata2019guiding} forces streaming and non-streaming CTC models to learn the same timings of spikes and then frame-level knowledge distillation is applied, which is also suitable in RNN-T~\cite{kurata2020knowledge}.

Pre-training is another alternative to improve the performance of streaming  CTC~\cite{yu2018multistage,zeyer2018improved,audhkhasi2017direct} and RNN-T~\cite{alex-rnnt-2013,google-sak-rnnt-2017,hu2020exploring,sougou-rnnt-2019}. For RNN-T, a pre-trained language model is used to initialize prediction network~\cite{google-sak-rnnt-2017}. It has been proven that RNN-T with encoder pre-trained using CTC loss gets better performance~\cite{alex-rnnt-2013, google-sak-rnnt-2017}, as RNN-T is regarded as an extension to CTC. Besides, encoder pre-training outperforms whole-network pre-training and frame-wise cross entropy (CE) criterion with external alignments outperforms CTC criterion~\cite{hu2020exploring}. 
Generally, frame-wise label for pre-training is produced by a HMM-based system~\cite{tan2021aispeech}, which is somewhat conflicting to the motivation of E2E ASR.

In this paper, we propose to pre-train streaming RNN-T's encoder with frame-wise label generated by an all-neural framework without using HMM-based system, named HMM-free pre-training. It needs a teacher CTC model and its alignment path to simulate the frame-wise label. To our best knowledge, it is the first time that CTC alignment is used to simulate HMM-based frame-wise label for CE pre-training.
%though CTC alignment can be used for other purpose \cite{moritz2019triggered}. 
It is hard to predict accurate token boundaries using CTC's spikes as they appear irregularly and tokens have different duration, which is alleviated by using soft labels.
Since CTC's modeling units can be characters or other units unrelated to acoustic pronunciation, the proposed method can be landed much faster in new languages where the pronunciation lexicon is not available.

The experiments are conducted on LibriSpeech~\cite{panayotov2015librispeech} and Multilingual LibriSpeech (MLS)~\cite{pratap2020mls} English corpora. Compared with the pre-training method using HMM-based  frame-wise alignments, the proposed method has comparable performance in both recognition accuracy and emission latency. Compared with the baseline without pre-training, word error rate (WER) of LSTM RNN-T model is reduced by relatively 5\%$\sim$11\%, and the emission latency is reduced by about 60 ms. 

The paper is organized as follows: Section \ref{rnnt} gives a brief introduction about RNN-T architecture. The proposed HMM-free pre-training procedure is discussed in Section \ref{sec_hmm_free}, followed by  experiments and discussions in Section \ref{src_exp}.

\section{RNN Transducer}
\label{rnnt}
\begin{figure}[t]
	\centering
	\includegraphics[width=\linewidth]{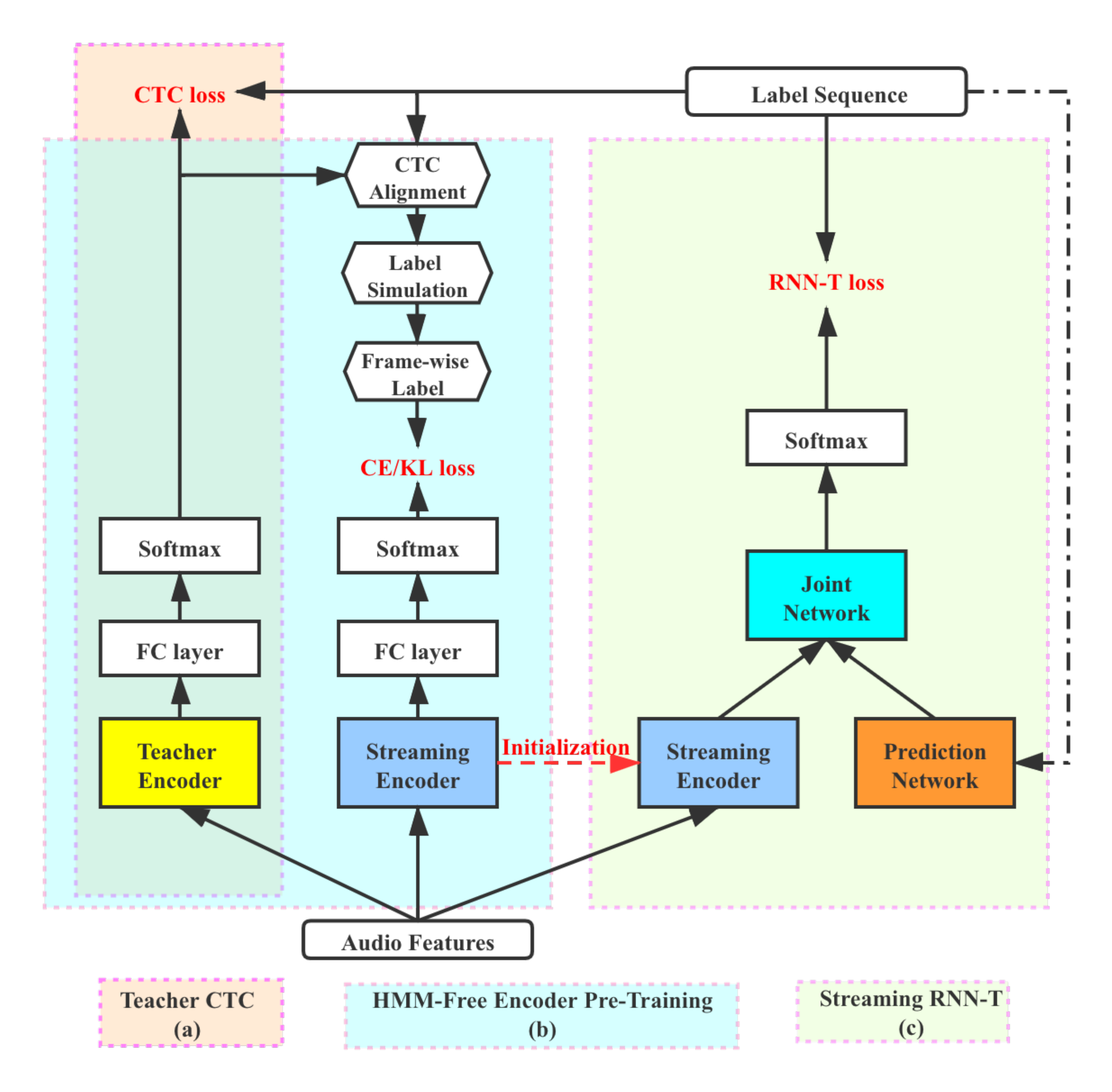}
	\caption{The model structures of teacher CTC, streaming encoder pre-training and streaming RNN-T. ``FC" layer denotes fully-connected layer.}
	\label{fig:model}
\end{figure}

As illustrated in Figure  \ref{fig:model}(c), RNN-T model consists of encoder, prediction network and joint network. The encoder converts the input $T$-frame acoustic features $\mathbf{x}=(x_1,\dots, x_T)$ to high-level representation $\mathbf{h}^{\text{enc}}=(h_1^{\text{enc}},\dots,h_T^{\text{enc}})$ by
\begin{equation}
	{h}_t^{\text{enc}}=f_{\text{enc}}(\mathbf{x}).
\end{equation}
%For streaming encoder, ${h}_t^{\text{enc}}$ is computed by using current and previous input features $\mathbf{x}_{1:t}$.

The prediction network receives historical non-blank labels $y_1, \dots, y_{u-1}$ as input to produce its output at step $u$
\begin{equation}
	h^{\text{pred}}_u=f_{\text{pred}}(\mathbf{y}_{1:u-1}).
\end{equation}

Then the logits over vocabulary at frame $t$ and step $u$ can be computed by the joint network
\begin{equation}
 	h^{\text{joint}}_{t,u}=f_{\text{joint}}({h}_t^{\text{enc}}, h^{\text{pred}}_u).
\end{equation}

Finally, the probability distribution over vocabulary at frame $t$ and step $u$ is calculated using a softmax layer. With forward-backward algorithm~\cite{graves2012sequence}, the sum probability $P(\mathbf{y}|\mathbf{x})$  of all alignment paths $\pi$ is adopted as the objective function
\begin{equation}
	\mathcal{L}=-\log(P(\mathbf{y}|\mathbf{x}))=-\log\sum_{\pi\in\Pi(\mathbf{y})}P(\mathbf{\pi}|\mathbf{x})
\end{equation}
where $\mathbf{y}=(y_1,\dots, y_U)$ is the label for training, $U$ is the number of target tokens, and $\Pi(\mathbf{y})$ is the alignment path sets.

During inference, the beam search algorithm described in \cite{graves2012sequence} is used with beam size 10. And neural network LM (NNLM) can be used to re-score the RNN-T's N-best hypotheses.

\section{HMM-Free Pre-Training}
\label{sec_hmm_free}
Encoder pre-training is important for RNN-T, especially for streaming RNN-T.
However, pre-training using sequence-level criterion like CTC often performs worse than frame-level criterion like CE~\cite{hu2020exploring}. This may be because CTC tends to predict blank symbol and leads to incorrect inference for RNN-T, while CE makes encoder a token classification model and thus RNN-T can learn more accurate token boundaries. 

Generally, frame-wise label for pre-training is produced by a HMM-based system~\cite{tan2021aispeech}, which depends a pronunciation model, and is somewhat conflicting to the motivation of E2E ASR. Hence, this work proposes a pre-training procedure by generating the frame-wise label using an all-neural framework, which is referred to as \texttt{HMM-free pre-training}. Besides, pre-training using HMM-based frame-wise label is denoted as \texttt{HMM-based pre-training}. 

The proposed method has three benefits.
Firstly, no pronunciation model is needed, which makes it friendly for new language expansion.
Besides, the encoder is pre-trained to learn fuzzy word boundaries generated by the teacher CTC model rather than the same timings of spikes as teacher model, which may provide more effective and supplementary information for RNN-T. Finally, it is noticed in the experiments that HMM-free pre-training helps reduce the emission latency.

\subsection{Training procedure}
As illustrated in Figure \ref{fig:model}, HMM-free pre-training procedure consists of three stages.
The first stage is to train a teacher CTC model, whose modeling units is the same as RNN-T's.

The next stage is to pre-train a streaming encoder using CE or Kullback–Leibler (KL) divergence loss. In this stage, frame-wise label for pre-training is simulated from teacher CTC model's alignment, where the later is computed using forward-backward algorithm~\cite{alex-ctc-2006}, and the detail of simulation will be described in Section \ref{sub_section_align}. 
Besides, only the streaming encoder is updated, while the parameters of teacher CTC model are fixed.

Finally, streaming RNN-T is trained with pre-trained encoder and random initialized prediction and joint network.

\subsection{Alignment Simulation}
\label{sub_section_align}

\subsubsection{Hard Label} 

\begin{figure}[t]
	\centering
	\includegraphics[width=\linewidth]{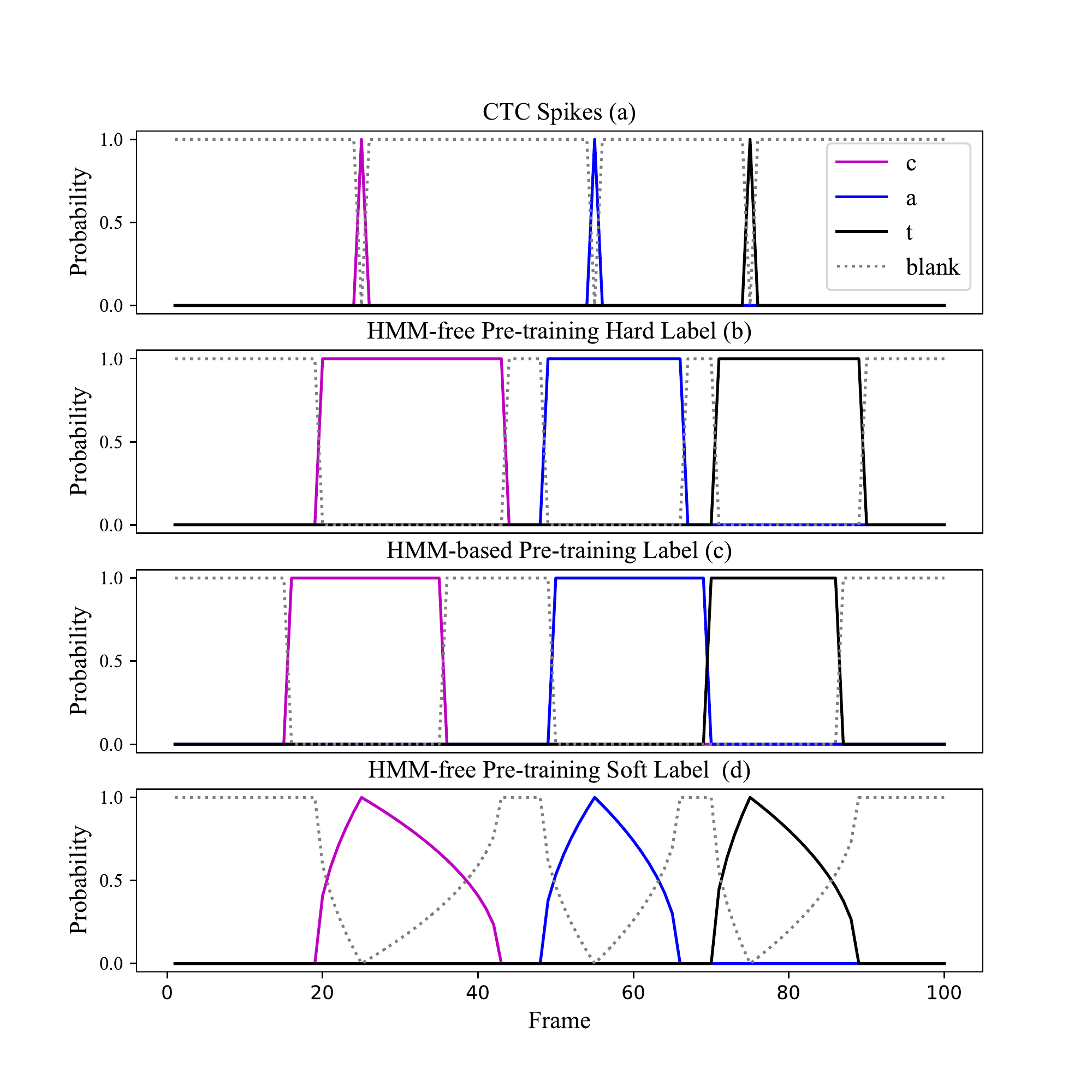}
	\caption{CTC's spikes, simulated ``hard" and ``soft" frame-wise labels, and HMM-based frame-wise label. The silence in HMM-based label is mapped as blank symbol.}
	\label{fig:alignment}
\end{figure}

As illustrated in Figure \ref{fig:alignment}(a) and  \ref{fig:alignment}(c), most of the frames are aligned to blank symbol in CTC alignment, while HMM-based alignment provides a chunk for each token across time axis. So we simply and naturally expand the spikes of teacher CTC model to their left/right blank frames by pre-defined left/right ratios to simulate HMM-based alignment, as shown in Figure \ref{fig:alignment}(b). We denote this type of frame-wise label as ``hard" label.

CE is used as the loss between streaming encoder's output and simulated frame-wise hard label
\begin{equation}
	\mathcal{L}_{\text{hard}} = -\frac{1}{T}\sum_{t=1}^T \log \hat{P}_{t}^{y_t}, 
\end{equation}
where $y_t$ is the hard label index at frame $t$, which can be blank or non-blank symbol, $\hat{P}_{t}=\text{Softmax}(\text{FC}(h_{t}^{\text{enc}}))$ is the predicted probability distribution at frame $t$ over all modeling units, and $\hat{P}_{t}^{y_t}$ is the predicted probability of label $y_t$.

\subsubsection{Soft Label}
Since the token's spikes in CTC don't appear in the exact middle of pronounced frames, it is hard to predict accurate token boundaries by simply expanding the spikes. As illustrated in Figure \ref{fig:alignment}(b) and  \ref{fig:alignment}(c), there are many mismatches near the token boundaries between hard label and HMM-based label.

To alleviate the incorrect predictions near boundaries, ``hard" label is improved to ``soft" label. This is achieved by giving probability smaller than 1.0 for frames away from the spike, with the remained left as the probability of blank, as shown in Figure \ref{fig:alignment}(d). For a frame $t$ with distance of $d$ frames to the spike, soft probability of non-blank token $y_t$ is calculated by
\begin{equation}
	P_{y_t}=\sqrt{\max(1.0 - d/W, 0)},
\end{equation}
where $W=\lfloor r*W_{blk}  \rfloor$ is the number of expanded frames, $W_{blk} $ is the number of blank frames in the left/right side, and $r$ is the expanding ratio, which can be different for left/right side. Besides, $1-P_{y_t}$ is left as the probability of blank symbol.

Similarly, the CE loss using ``soft" label is calculated  by
\begin{equation}
	\mathcal{L}_{\text{soft}} = -\frac{1}{T}\sum_{t=1}^T\left(P_{y_t}\log \hat{P}_{t}^{y_t} +(1-P_{y_t})\log \hat{P}_{t}^{blk}\right),
\end{equation}
%$P_{y_t}$ is the soft probability of non-blank symbol $y_t$ at frame $t$, and it is 0 for the blank frames. 
where $\hat{P}_{t}^{y_t}$ and $\hat{P}_{t}^{blk}$ are the predicted probabilities of non-blank symbol $y_t$ and blank symbol at frame $t$ respectively. This is the same as KL divergence, because target distribution only has positive probabilities over non-blank symbol  $y_t$ and blank symbol at frame $t$, and the entropy of target probability distribution doesn't have gradients.

\section{Experiments}
\label{src_exp}
\subsection{Experimental Setup}
The experiments are conducted using LibriSpeech and MLS English corpora, which have 960-hour and 44.5k-hour audios for training respectively. The WER is evaluated on LibriSpeech \texttt{test-clean/test-other} and MLS \texttt{dev/test} test sets respectively. The models are trained with sub-word units, with about 5k/10k units for Librispeech/MLS English respectively.

The 80-dim filter-bank features are stacked by 4 frames and without skipping, and then fed into the encoder. Besides, all models are trained with Spec-Augment~\cite{park2019specaugment}.
In this work, non-streaming Bi-directional LSTM (BLSTM) is used as encoder of teacher CTC model, and LSTM is used as streaming encoder.
The LSTM encoder contains 8 layers with 1024 units and a projection layer with 640 units. There is a residual connection between the input and output of each layer.
Similar to~\cite{google-device-rnnt-2019}, two time reduction layers are added after the first and second layer to down-sample the frame rate to 4. The BLSTM encoder has the same structure as LSTM and the cell size is changed to 512. For RNN-T's prediction network, 2-layer LSTM without projection layer is adopted. RNN-T's joint network is a feed-forward layer. All the neural networks are trained using PyTorch~\cite{paszke2019pytorch}.

Emission latency is used to evaluate streaming RNN-T's recognition latency. Similar to the word-level latency~\cite{wang2020low}, for each word, it is defined as the difference between ground truth end time and recognized end time. The ground truth end time is generated from HMM-based forced alignment. The recognized time is defined as the frame when RNN-T recognizing the word's last sub-word. The emission latency can be negative as the word can be recognized before its ground truth end time.

\subsection{Librispeech Baseline}
Firstly, LSTM RNN-T and BLSTM RNN-T are trained from scratch on LibriSpeech task. As presented in Table \ref{exp_baseline}, without NNLM, BLSTM RNN-T achieves WER 4.02\%/9.67\% on test-clean/test-other. While LSTM RNN-T achieves 6.03\%/14.23\%, and it is about 50\%/47\% relatively worse respectively.

Besides, a NNLM consisting of 2-layer LSTM is trained using the same sub-word units on Librispeech language model corpus\footnote{\url{http://openslr.org/11}}. 
We apply LM re-scoring to the N-best hypotheses during beam search.
As shown in Table \ref{exp_baseline}, NNLM brings a relative gain of 7\%$\sim$16\% on two test sets. It should be noted that the NNLM used in this work is trained with sub-word units, while it may get better performance when trained using words.

Also, the result of LSTM RNN-T with encoder pre-trained using CTC criterion is presented in Table \ref{exp_baseline}, which is just slightly better than the baseline trained from scratch on the test-clean test set.

Finally, for LSTM RNN-T, two additional baselines with encoder pre-trained using HMM-based frame-wise labels are also trained. Kaldi~\cite{povey2011kaldi} is used to train a HMM-based system and generate frame-wise phone and word labels on LibriSpeech task. 
%It should be noted that although a deep neural network-HMM (DNN-HMM) system can be trained to generate more accurate frame-wise alignments, we leave this in the future work. 
%as the label used for training DNN-HMM is also generated by GMM-HMM system.
The phone labels are used to pre-train the encoder directly. 
The word labels are firstly converted to sub-word labels by assigning the duration of each word to its sub-word units on average and then used for pre-training. The results are denoted as \texttt{HMM-based pre-training (phone/sub-word)} in Table \ref{exp_baseline}. LSTM RNN-T pre-trained using frame-wise sub-word labels outperforms the one using phone labels. Compared with LSTM baseline without pre-training, there are relatively 11\% and 5\% gains on test-clean and test-other respectively. With NNLM, the relative gap between LSTM and BLSTM RNN-T is reduced from 47\%$\sim$50\% to 31\% $\sim$ 36\%.

\begin{table}[th]
	\caption{WER(\%) of LSTM/BLSTM RNN-T baselines on LibriSpeech task.}
	\label{exp_baseline}
	\centering
	\begin{tabular}{l|c|c|c}
		\toprule
		model & NNLM & \makecell[c]{test\\clean} & \makecell[c]{test\\other} \\
		\midrule
		\multirow{2}{*}{BLSTM}   &  no & 4.02 & 9.67  \\
		&  yes & \textbf{3.56} & \textbf{8.98}  \\
		\midrule
		\multirow{2}{*}{LSTM}    &    no & 6.03  & 14.23    \\
		&    yes & 5.18   & 12.99   \\
		\midrule
		\multirow{2}{*}{\makecell[c]{+ CTC pre-training}}& no &  5.81 &  14.40  \\
		& yes & 4.96 &  12.95 \\
		\midrule
		\multirow{2}{*}{\makecell[c]{+ HMM-based pre-training\\(phone)}}& no &  5.69 & 13.82  \\
		& yes & 4.85 & 12.46 \\
		\midrule
		\multirow{2}{*}{\makecell[c]{+ HMM-based pre-training\\(sub-word)}}  & no & 5.35 & 13.52   \\
		&   yes & \textbf{4.67} & \textbf{12.21} \\
		\bottomrule
	\end{tabular}
\end{table}

\subsection{Performance of HMM-Free Pre-Training}
\subsubsection{Hard Label and Soft Label}
We firstly evaluated the performance of proposed HMM-free pre-training with hard labels. For HMM-free pre-training, we sweep the parameters of left/right ratios and we find 0.2/0.6 works well, which is adopted as default configuration in the following experiments. It is noticed that the sum of left/right ratios is smaller than 1.0, where the remained frames are left as blanks to make RNN-T model learn token boundaries more easily.

As illustrated in the Table \ref{exp_librispeech}, results using hard labels are better than the baseline without any pre-training by relatively 5\%$\sim$7\%. Compared to the baseline pre-trained using HMM-based sub-word labels, WER on test-clean is slightly worse by relatively 4\%. When NNLM is adopted, the gap on test-clean goes to 2.6\%, and HMM-free pre-training is even slightly better than HMM-based pre-training on test-other by relatively 1.6\%.
 
\begin{table}[th]
	\caption{WER(\%) of LSTM RNN-T using HMM-free pre-training with hard/soft label on LibriSpeech task.}
	\label{exp_librispeech}
	\centering
	\begin{tabular}{l|c|c|c}
		\toprule
		model & NNLM & \makecell[c]{test\\clean} & \makecell[c]{test\\other} \\
		\midrule
		\multirow{2}{*}{LSTM} & no   & 6.03 & 14.23  \\
		& yes  &    5.18   & 12.99   \\
		\midrule
		\multirow{2}{*}{\makecell[c]{+ HMM-based pre-training\\(sub-word)}} & no & 5.35 & 13.52  \\
		& yes  & \textbf{4.67}& 12.21\\
		\midrule
		\multirow{2}{*}{\makecell[c]{+ HMM-free pre-training \\(hard)}} & no & 5.57 & 13.51   \\
		& yes   & 4.79 &  \textbf{12.01}     \\
		\midrule
		\multirow{2}{*}{\makecell[c]{+ HMM-free pre-training \\(soft)}}  & no & 5.40   & 13.44  \\
		& yes      &  \textbf{4.62} & \textbf{12.07}         \\
		\bottomrule
	\end{tabular}
\end{table}

Also, we evaluate the performance of HMM-free pre-training with soft label. As illustrated in the last row in Table \ref{exp_librispeech}, HMM-free pre-training with soft label outperforms HMM-based pre-training slightly. It outperforms the baseline trained from scratch by relatively 7\%$\sim$11\%.

\subsubsection{Larger Data Set: MLS English Corpus}
\begin{table}[th]
	\caption{WER(\%) of LSTM RNN-T with HMM-free pre-training with hard/soft label on MLS English task.}
	\label{exp_mls}
	\centering
	\begin{tabular}{l|c|c|c}
		\toprule
		model & NNLM & dev & test \\
		\midrule
		\multirow{2}{*}{LSTM}  &  no & 11.91 & 13.16  \\
		& yes & 10.27 & 11.46 \\
		\midrule
		\multirow{2}{*}{\makecell[c]{+ HMM-free pre-training \\(hard)}} & no & 10.96 & 12.27         \\
		& yes & \textbf{9.62} & \textbf{10.73} \\
		\midrule
		\multirow{2}{*}{\makecell[c]{+ HMM-free pre-training \\(soft)}}  & no & 10.48 & 11.88  \\
		&yes & \textbf{9.47} & \textbf{10.50} \\
		\bottomrule
	\end{tabular}
\end{table}

In this section, we evaluate the performance of HMM-free pre-training on MLS English task. We did not conduct HMM-based experiments as the generation of HMM-based alignments is time-consuming. Similarly, a 2-layer LSTM NNLM is also trained with the training transcripts.

The results are reported in Table \ref{exp_mls}. When using hard label, HMM-free pre-training outperforms the baseline trained from scratch with relatively 6\% WER reduction on dev/test set. Besides, NNLM brings further relatively 13\% WER reduction. Finally, compared to hard label, soft label brings additional 2\% relative WER reduction.

\subsection{Emission Latency}
Finally, we discuss the latency issue of LSTM RNN-T models with and without pre-training, and the models pre-trained using different methods as well. We only select the utterances recognized exactly the same as their references to compute the emission latency. The median and 90th percentile of emission latency are evaluated on LibriSpeech test-clean test set, denoted as \texttt{EL@50} and \texttt{EL@90} respectively.

The latency results are summarized in Table \ref{exp_latency}.
Compared with the baseline without pre-training, LSTM RNN-T models using HMM-based and HMM-free pre-training with hard label reduce \texttt{EL@50} and \texttt{EL@90} by 60$\sim$70 ms, while pre-training encoder using CTC criterion has no benefit.
This is because the frame-wise alignments and criterion help the model learn more accurate token boundaries.
Also, it is noticed that HMM-free pre-training with soft label has the same latency as that with hard label.

\begin{table}[th]
  \caption{The emission latency on LibriSpeech test-clean test set.}
  \label{exp_latency}
  \centering
  \begin{tabular}{l|c|c}
    \toprule
    model &   \texttt{EL@50} &  \texttt{EL@90}  \\
    \midrule
    LSTM  &     230ms & 330ms      \\
    + CTC pre-training & 220ms &330ms \\
    + HMM-based pre-training & 160ms &	270ms \\
    + HMM-free pre-training (hard) &170ms & 270ms  \\
    + HMM-free pre-training (soft)&   170ms   &  270ms   \\
    \bottomrule
  \end{tabular}
\end{table}

\section{Conclusions}
\label{sec_con}
In this paper, the proposed HMM-free pre-training procedure with hard/soft label is evaluated on LibriSpeech and MLS English tasks. The procedure is an all-neural framework without using HMM-based models, and has comparable or even slightly better performance as HMM-based pre-training method in both recognition accuracy and emission latency. On Librispeech task, compared with the LSTM RNN-T trained from scratch, HMM-free pre-training gives 5\%$\sim$11\% relative improvements on recognition accuracy, and it helps reduce the emission latency by 60 ms. Besides, HMM-free pre-training with soft label also slightly outperforms that with hard label. Though the procedure is evaluated with LSTM/BLSTM models, it can also be applied to other types of non-streaming and streaming models.

%\section{Acknowledgements}
%The authors thanks 

\bibliographystyle{IEEEtran}

\bibliography{mybib}

\end{document}